# Impacts of transport development on residence choice of renter households: An agent-based evaluation

**Ali Shirzadi Babakan and Mohammad Taleai**

**Abstract**

Because of improving accessibility, transport developments play an important role on residence choice of renter households. In this paper, an agent-based model is developed to investigate impacts of different transport developments on residence choice of renter households in Tehran, the capital of Iran. In the proposed model, renter households are considered as agents who make a multi-objective decision and compete with each other to rent a preferred residential zone. Then, three transport development scenarios including construction a new highway, subway and bus rapid transit (BRT) line are simulated and resulting changes in residence choice of agents are evaluated. Results show that transport development scenarios significantly affect residence choice behavior of different socio-economic categories of renter households and lead to considerable changes in the residential demand, composition of residents, mean income level and mean car ownership in their vicinities.

## 1. Introduction

A mutual relationship between land uses and transportation has been shown by many researchers (Iacono et al. , 2008, Waddell, 2011, Wegener, 2014). Development and changes of land uses influence the transportation system and on the other hand, development and changes of the transportation system affect land use patterns. For example, development of a new highway leads to changes in surrounding land uses and reciprocally by altering travel demands, affects the transportation system. In order to study this relationship, various land use-transportation models have been developed using various mathematical, statistical, heuristic, and microsimulation methods. Detailed reviews of these models are available in (Iacono et al. , 2008, Waddell, 2011, Wegener, 2014).

One of the most essential parts of land use-transport models is modeling of residence choice of households. This modeling includes a complex decision-making process in which households select their residence by considering various criteria. Households often choose their residence according to their socio-economic characteristics (e.g. income, the number of members, and the number of owned cars), neighborhood characteristics (e.g. housing price and accessibility to various services and opportunities), and accessibility to pre-specified destinations such as workplaces (Jordan et al. , 2012, Jun et al. , 2013, Wang and Li, 2006, Wu et al. , 2013, Yang et al. , 2013). Therefore, transportation system has a major role in this decision making, because it directly affects the accessibility to various services and pre-specified destinations.

The impacts of transport developments on the housing market and residential decisions of households have been studied by many researchers. Pagliara et al. (2010) applied a bid-choice model and an hedonic model to study the impacts of different transport policies such as road user charging, changes to fuel duties and the provision of light rapid transit systems on residential location decisions, housing occupancy rates and housing prices in the Greater Oxford area. Their research shows that transport policies have significant impacts on the housing market. For example, the road user charging might reduce the average housing price about 2%. Also, development of a new public transport system might increase the housing price by around 3% on average. Eliasson (2010) attempted to examine the influence of accessibility on residential location choice of households using TILT, a land use-transportation model for the Stockholm region. He found that the attractiveness of a location increases with the improvement of accessibilities to workplaces and different types of service. Cervero and Kang (2011) investigated the impacts of converting regular bus operations to bus rapid transit (BRT) on land use changes and land values in Seoul, Korea. Their findings showed that development of BRT leads to convert single-family residences to higher density apartments. Also an increase of more than 10% for values of residential land uses within 300 m of BRT stops is revealed. Calvo et al. (2013) evaluated the impacts of development of the subway system on population and land uses in Madrid, Spain. They found that considerable residential developments and population growth are observed in vicinities of the new subway stations. Also, urbanization and population settlement are significantly more dynamic in proximities of the subway developments. They suggested that the population density reduces by increasing distance from the new subway stations. Mathur and Ferrell (2013) estimated the impact of a light rail line on single-family home prices using a hedonic regression model in San Jose, CA. They found that the average home sale price within 1/8 mile of the light rail stations increases about 3.2% for every 50% reduction in the distance between the home and stations. In addition, the average housing price within 1/8 mile of the light rail line is 18.5% higher than the average price within distances of more than 1/8 mile from the light rail line. Zhang et al. (2014) examined the impacts of different transit systems including bus rapid transit (BRT), light rail transit (LRT) and metro rail transit (MRT) on the price of residential properties using hedonic price models. Their research showed that transit impacts on the housing market extend to 1 mile for MRT, 0.5 mile for LRT and are discernable for BRT stations. Also, the value of residential properties increases 39.41 and 17.57 USD/m2 for every 100 m closer to the stations of MRT and LRT, respectively. Hamersma et al. (2015) studied the trade-off between nuisances and accessibility in residential moving intentions of people living near highways using structural equation modeling in the Netherlands. Their study investigated that highway usage and other residential characteristics including satisfaction with buildings, traffic safety, and amount of green spaces may compensate perceived highway nuisances. Also, some groups of residents such as home owners are less tended to move without considering their residential satisfaction. However, a majority of these studies are aggregate and insensitive to the behavior of individual households (Benenson, 2004). This issue has prompted researchers to use disaggregate models which can represent decision of individual households.

One of the most applicable methods which has attracted the attention of researchers for disaggregated modeling of residence choice of households is agent-based modeling. Agent-based model is a 'bottom-up' approach in which behaviors and interactions of agents are characterized and used to produce different aggregated results. Agents are autonomous entities who can perceive their environment, move through space and time, and take actions based on their objectives (Crooks and Heppenstall, 2012). Agent-based models have many capabilities for studying residence choice of households including clear representation and interpretation of behavior of households (Barros, 2004), better understanding of quantitative and qualitative changes in the urban system (Benenson, 2004), reduction of the computational complexity, explicit representation of heterogeneity and interdependencies among households and their environment, considering various conditions and constraints on households and flexible aggregation of results (Barros, 2004, Hunt, 2002). Therefore, a number of researchers have used agent-based models to study residence choice behavior of households (e.g. Benenson, 2004, Devisch et al. , 2009, Ettema, 2011, Gaube and Remesch, 2013, Haase et al. , 2010, Jordan et al. , 2012). A detailed review of these models can be found in (Huang et al. , 2014). In addition to these models, a new generation of land use-transportation models including RAMBLAS (Veldhuisen et al. , 2000), UrbanSim (Waddell et al. , 2003) and ILUTE (Salvini and Miller, 2005) have been developed using agent-based modeling. These comprehensive urban models have been applied by several researchers (e.g. (Farooq and Miller, 2012, Kakaraparthi and Kockelman, 2011, Kryvobokov et al. , 2015, Veldhuisen et al. , 2005, Waddell et al. , 2007)) to model the interactions between land uses and transportation system and also to evaluate impacts of various land use and transport "what if" scenarios on housing market and residential decisions of households in many metropolitan regions.

In this paper, an agent-based model is used to investigate role of different transport development scenarios in the residence choice process of renter households. Majority of previous studies have used discrete choice models based on the random utility maximization theory for determination of households' residence. In this theory, households calculate utility of a finite number of well-identified options and select the one with the maximum utility. But, in this paper, a multi-objective decision making method and a heuristic competition method are used to determine renter households' residence. In the proposed agent-based model, renter households are considered as agents who have different criteria and preferences depending on their socio-economic and demographic characteristics. They search among residential zones and select their appropriate residential options according to their criteria and preferences. Subsequently, they compete with each other to rent a residence among their preferred options in different time periods. Finally, three transport development scenarios are simulated and changes in residence choice of agents are evaluated.

The proposed model has been applied in Tehran, the capital of Iran. Tehran with an area of about 750 square kilometers and a population of about 8.3 million is one of the largest and the most populated capitals in the world (Tehran Municipality, 2013b). This metropolis has been divided into 560 traffic analysis zones (TAZs) which have been used as the spatial units in this research. Tehran consists of a wide transit system including highway, subway, bus and BRT networks. This metropolis includes a large number of highways with the length of about 550 kilometers, a subway network composed of 4 active intraurban lines with the length of about 125 kilometers and a bus network comprised of 250 lines with the length of about 3000 kilometers. In addition to these networks, 6 bus rapid transit (BRT) lines with the length of about 102 kilometers are operational in Tehran. Although the length of BRT lines is about 3% of the total length of the bus lines, about 40% of bus passengers are transported with these lines (Tehran Municipality, 2013a). This shows the popularity and effectiveness of BRT network in transit system of Tehran. In recent decades, insufficient development of the public transit on one side and low prices of fuel on the other side have resulted in different problems including traffic congestion and air pollution in Tehran. Therefore, urban planners and policymakers attempt to resolve these problems by developing various urban plans such as improvement of public transit and highway networks in Tehran. However, these plans have side effects on the other urban activities including the residence choice process of households. In this paper, it is attempted to study these effects.

The paper is structured in 6 sections. The proposed agent-based model is described in detail in section 2. Sections 3 and 4 highlight implementation and validation procedures of the proposed model in Tehran metropolis. In section 5, different transport development scenarios are simulated and their effects on the residence choice of households are evaluated. Finally, conclusions of the research are presented in section 6.

## 2. Residence choice model

In the proposed model, renter households are considered as agents who look for an appropriate residence among residential zones. In reality, renters are unable to search all zones, but they usually select some zones to search a residence among them. The same procedure is used in this model. At the first step, a synthetic population of renter households is generated using the Monte Carlo simulation. Then, agents select their residential options according to their socio-economic characteristics and desired objectives. In fact, they face with a multi-objective decision making problem, because some of their objectives (e.g. minimizing housing rent and maximizing accessibility to different services) conflict with each other. A constrained NSGA-II algorithm is developed to find the optimal residential options with respect to their objectives. Finally, they compete with each other to rent a residence among their desired residential options. The main parts of the proposed model are described below.

## 2.1. Population synthesis

In this section, the available aggregated demographic data in each zone (Tehran Municipality, 2013b) are used as references to generate renter households (agents) with different attributes. For this purpose, a sequential approach using the Monte Carlo simulation is developed in which attributes of agents are simulated in each step based on the determined attributes in the previous step. In this approach, at first, monthly income, number and age of the members of agents are randomly simulated in each zone. Then, number of owned cars and employees, and required residential area of agents are randomly generated according to the aforementioned attributes. In the next step, preferences and criteria of each agent for selecting a residence are randomly generated based on the previously determined attributes. Finally, the existing information of employment distribution in different zones and the general pattern of distances between workplace and residence of employees (TCTTS, 2012) are used to define the workplace of employed members of agents.

Data provided by a field survey of the stated preferences of residence choice of sample renter households in Tehran are used to simulate the residential criteria and preferences of agents. These data were collected from 480 renter households with different socioeconomic and cultural characteristics by using questionnaires in April 2013. They were requested to state their residential criteria and preferences in the scale range of 0 (unimportant) to 9 (very important). In Table 1, these sample households have been classified by the household size, average monthly income and number of owned cars. Then, in each class, the percentage of households who stated a preference number of greater than 4 to each criterion has been specified.

Table 1: A summary of preferred residential criteria of sample renter households in Tehran

| Attribute | Class | Percentage | Housing Rent | Distance from the Workplace | Distance from the Former Residence | Air Pollution | Noise Pollution | Accessibility to retail stores | Accessibility to Educational centers | Accessibility to Green and Recreational spaces | Accessibility to health Centers | Accessibility to Cultural Centers | Traffic Restrictions | Accessibility to Highway network | Accessibility to Subway Stations | Accessibility to Bus Stops |
|---|---|---|---|---|---|---|---|---|---|---|---|---|---|---|---|---|
| Size | Single | 6.9 | 100 | 92.9 | 46.4 | 50.0 | 64.3 | 46.4 | 17.9 | 7.1 | 3.6 | 10.7 | 67.9 | 78.6 | 35.7 | 17.9 |
| | Couple | 35.6 | 100 | 90.7 | 52.6 | 63.9 | 74.2 | 61.9 | 10.3 | 50.5 | 7.2 | 7.2 | 64.9 | 73.2 | 48.5 | 20.6 |
| | 3-4 | 44.4 | 100 | 84.2 | 62.6 | 61.4 | 69.6 | 62.6 | 71.3 | 57.3 | 12.3 | 8.8 | 63.7 | 71.3 | 53.2 | 22.8 |
| | > 4 | 13.1 | 100 | 79.4 | 61.8 | 64.7 | 70.6 | 55.9 | 88.2 | 64.7 | 14.7 | 5.9 | 67.6 | 73.5 | 52.9 | 20.6 |
| Average Monthly income (USD) | < 500 | 29.4 | 100 | 92.8 | 68.1 | 40.6 | 47.8 | 47.8 | 42.0 | 34.8 | 7.2 | 5.8 | 46.4 | 58.0 | 58.0 | 29.0 |
| | 500-1000 | 51.9 | 100 | 85.8 | 56.4 | 64.2 | 73.5 | 64.7 | 58.3 | 53.9 | 12.3 | 8.8 | 64.7 | 70.6 | 49.5 | 21.1 |
| | > 1000 | 18.8 | 100 | 80.7 | 52.6 | 77.2 | 87.7 | 59.6 | 33.3 | 64.9 | 7.0 | 8.8 | 87.7 | 98.2 | 43.9 | 14.0 |
| Number of Cars | 0 | 11.7 | 100 | 97.1 | 55.9 | 38.2 | 52.9 | 64.7 | 55.9 | 44.1 | 14.7 | 8.8 | 5.9 | 14.7 | 88.2 | 73.5 |
| | 1 | 66.5 | 100 | 86.1 | 58.4 | 62.8 | 70.6 | 63.2 | 55.8 | 55.8 | 10.8 | 8.2 | 64.5 | 73.6 | 55.4 | 18.2 |
| | > 1 | 21.9 | 100 | 81.5 | 58.5 | 69.2 | 80.0 | 47.7 | 29.2 | 41.5 | 6.2 | 7.7 | 96.9 | 100 | 12.3 | 6.2 |
| Total | | 100 | 100 | 86.4 | 58.2 | 61.5 | 70.6 | 60.3 | 50.6 | 51.8 | 10.3 | 8.2 | 64.8 | 72.7 | 50.3 | 21.5 |

## 2.2. Determination of residential options of agents

In this section, agents independently choose their desirable residential zones without considering choices of other agents and residential capacity of zones. They select their residential options according to various criteria including housing rent, distance from their former residence, accessibility to public and transport services, environmental pollutions, traffic restrictions, and distance from their workplaces. These are the most important criteria which are derived from the survey of stated preferences of residence choice of renter households in Tehran. Agents select maximum number of ten residential options using the non-dominated sorting genetic algorithm II (NSGA-II). NSGA-II proposed by Deb et al. (2002) is one of the fast and most efficient multi-objective evolutionary algorithms which has been successfully applied to solve a wide range of multi-objective optimization problems. The main steps of this algorithm are outlined in Table 2 (For more details refer to (Deb et al., 2002)).

Table 2: The main steps of NSGA-II

| NSGA-II Framework |
|---|
| 1. Random generation of an initial parent population ($P_0$) of size $N$. |
| 2. Generation of an offspring population ($Q_0$) of size $N$ using common genetic operators |
| 3. $P_t=P_0$ and $Q_t=Q_0$. |
| 4. Generation of a combined population ($R_t$) of size $2N$ by $R_t=P_t \cup Q_t$. |
| 5. Generation of non-dominated fronts ($F_i$) from solutions of $R_t$ based on the definition of non-domination. The first front contains solutions which do not dominate each other and dominate all the other solutions. Similarly, this process is continued until all remaining solutions of $R_t$ are assigned to a front. |
| 6. Generation of $P_{t+1}$ of size $N$ from the solutions assigning to the first (best) fronts of $R_t$. |
| 7. Generation of $Q_{t+1}$ of size $N$ by applying binary tournament selection, crossover and mutation operators on $P_{t+1}$. In the binary tournament selection, the solution with lower domination rank or higher crowding distance is selected as the winner. The crowding distance of an individual solution is the perimeter of cuboid formed by its nearest neighboring solutions in the objective space that shows the density of solutions surrounding that solution. |
| 8. Repetition of steps 4 through 7 until the convergence criterion is met. |

In this paper, the following objective functions are used in NSGA-II algorithm to find optimal residential options of agents. It should be noted that depending on their residential criteria and preferences determined by the Monte Carlo simulation, agents consider one, some or all of the following objectives in their residence choice process.

*Housing rent:* All agents attempt to find a residence with minimum housing rent and in compatible with their income level using Eq. (1).

$$\text{Eq. (1)} \quad \min_{\substack{\forall a \in S_a \\ \forall i \in S_z}} (A_a * R_i) \quad (where: L_a^{min} * I_a \leq A_a * R_i \leq L_a^{max} * I_a)$$

where:
$S_a$ is the set of agents who the criterion is important for them;
$S_z$ is the set of zones;
$A_a$ is the required residential area of agent (*a*);
$R_i$ is the average housing rent per square meter in zone (*i*);
$I_a$ is the monthly income of agent (*a*);
$L_a^{min}$, $L_a^{max}$ are the limits on the monthly income of agent (*a*) for renting a residence. These limits are determined in the Monte Carlo simulation and restrict the search space of agents to zones in which they afford to rent a residence, where $0 \leq L_a^{min}$, $L_a^{max} \leq 1$;

*Accessibility to public services*: Almost all agents attempt to find a residence with maximum accessibility to their preferred public services including educational, retail, green and recreational, cultural or health services using Eq. (2) developed by Tsou et al. (2005).

$$\text{Eq. (2)} \quad \max_{\substack{\forall a \in S_a \\ \forall i \in S_z}} \left( \sum_s \sum_{j_s} p_a^s * w_{j_s} * d_{ij_s}^{-2} \right)$$

where:
*s* is the type of public services;
$j_s$ is the case (*j*) of the public service (*s*);
$p_a^s$ is the preference of agent (*a*) to the public service (*s*), where $\sum p_a^s = 1$;
$w_{j_s}$ is the relative effect of ($j_s$) which is calculated by $w_{j_s} = A_{j_s}/\max(A_{j_s})$, where ($A_{j_s}$) is the area of ($j_s$);
$d_{ij_s}$ is the distance between zone (*i*) and ($j_s$);

***Accessibility to transport services***: Almost all agents attempt to find a residence with maximum accessibility to their preferred transport services including bus stops, subway stations or highways using Eq. (3) developed by Currie (2010).

$$\text{Eq. (3)} \quad \max_{\substack{\forall a \in S_a \\ \forall i \in S_z}} \left( \sum_t \sum_{j_t} \frac{A_{j_t}}{A_i} * p_a^t \right)$$

where:
*t* is the type of transportation services;
$j_t$ is the case (*j*) of the transportation service (*t*);
$A_{j_t}$ is the area of service range of ($j_t$) which is inside zone (*i*);
$A_i$ is the area of zone (*i*);
$p_a^t$ is the preference of agent (*a*) to the transportation service (*t*), where $\sum p_a^t = 1$;

***Distance from the workplace:*** A great number of agents attempt to find a residence with minimum distance from their workplaces using Eq. (4).

$$\text{Eq. (4):} \quad \min_{\substack{\forall a \in S_a \\ \forall i, w_{e_a} \in S_z}} \left( \sum_{e=1}^n d_{iw_{e_a}} \right)$$

where:
$d_{iw_{e_a}}$ is the distance between zone (*i*) and the workplace of agent's employee ($w_{e_a}$);
*n* is the number of employees of agent (*a*);

***Distance from the former residence:*** Because of different reasons such as familiarity and dependency to the former residential area, many agents attempt to find a residence in proximity of their former residence using Eq. (5).

$$\text{Eq. (5)} \quad \min_{\substack{\forall a \in S_a \\ \forall i, f_a \in S_z}} (d_{if_a})$$

where:
$d_{if_a}$ is the distance between zone (*i*) and the former residence of agent (*f$_a$*);

***Air and noise pollutions:*** Many agents attempt to find zones with minimum air and noise pollutions using Eqs. (6) and (7).

$$\text{Eq. (6)} \quad \min_{\substack{\forall a \in S_a \\ \forall i \in S_z}} (P_i) \quad (where: \{P_i \leq 2\}, \forall a \in S_p)$$

$$\text{Eq. (7)} \quad \min_{\substack{\forall a \in S_a \\ \forall i \in S_z}} (N_i) \quad (where: \{N_i \leq 2\}, \forall a \in S_n)$$

where:
$P_i$, $N_i$ are the annual average of air and noise pollutions in zone (*i*); where they are classified in five categories varying from clean (0) to highly polluted (4);
$S_p$, $S_n$ are the sets of agents who air and noise pollutions are very important for them and their residential options are selected among clean (0) to medium (2) pollution categories;

***Traffic restrictions:*** Many agents attempt to find zones with minimum traffic restrictions on private cars including restrictions on all or odd-even private cars using Eq. (8).

$$\text{Eq. (8)} \quad \min_{\substack{\forall a \in S_a \\ \forall i \in S_z}} (T_i) \quad (where: \{T_i = 0\}, \forall a \in S_t)$$

where:
$T_i$ is the traffic restriction in zone (*i*) classified in 3 categories including (0) for no traffic restrictions; (1) for odd-even car restriction; and (2) for restriction on all private cars;

$S_t$ is the set of agents who traffic restrictions are very important for them and their residential options are selected among zones with no traffic restrictions.

### 2.3. Determination of final residence of agents

In this step, agents compete with each other in different months of the year to select their final residence among their preferred residential options. For this purpose, the relocation time of each agent is randomly assigned to a month of the year in accordance with the available statistical information of volume of relocation in different months of the year. Also, residential capacities of zones are limited in each month and computed by Eq. (9).

$$\text{Eq. (9):} \qquad C_i^t = N_i^t + H^t \frac{A_i^r}{\sum_i A_i^r}$$

where:
$C_i^t$ is the residential capacity of zone (*i*) at month (*t*);
$A_i^r$ is the residential area in zone (*i*);
$N_i^t$ is the number of agents whose former residences are located in zone (*i*) and want to change their residence at month (*t*);
$H^t$ is the number of new dwellings entered into the rental housing market at month (*t*) which is estimated using the available statistical information.

In each month, some agents search among their residential options according to the closeness of their options to their former residence or workplace and choose the first option having the capacity. If some agents simultaneously select an option which has not enough capacity for all of them, they compete with each other to reside in that zone. In this competition, agents with fewer members (except singles), higher income level and with no child will have a higher chance to succeed respectively, because most owners in Tehran prefer to rent their properties to these types of renters. In case of the similarity of qualifications of competitors, winners are selected randomly and defeated agents have to compete on their next options. This process is continued until either all agents reside in one of their options or the evaluation of all options of defeated agents is finished. Agents who are not able to reside in any of their residential options will have the opportunity to compete again with other agents in the next month. If they cannot reside in a zone in the next month, no residence would be considered for them.

### 3. Implementation

For simulating residence choice of renter households in Tehran, 50,000 agents are generated using the Monte Carlo simulation and their residence are determined by the developed NSGA-II and competition modules in MATLAB 7.9 software. Spatial distribution of residential options and final residence of agents have been shown in Figures 1 and 2, respectively. As seen in these Figures, The density of residential options and residents in central areas is higher than the other areas. This suggests that the central areas of Tehran generally are more attractive for renter households due to some reasons such as low housing rents and high accessibilities to public transport services, various public facilities and employment opportunities.

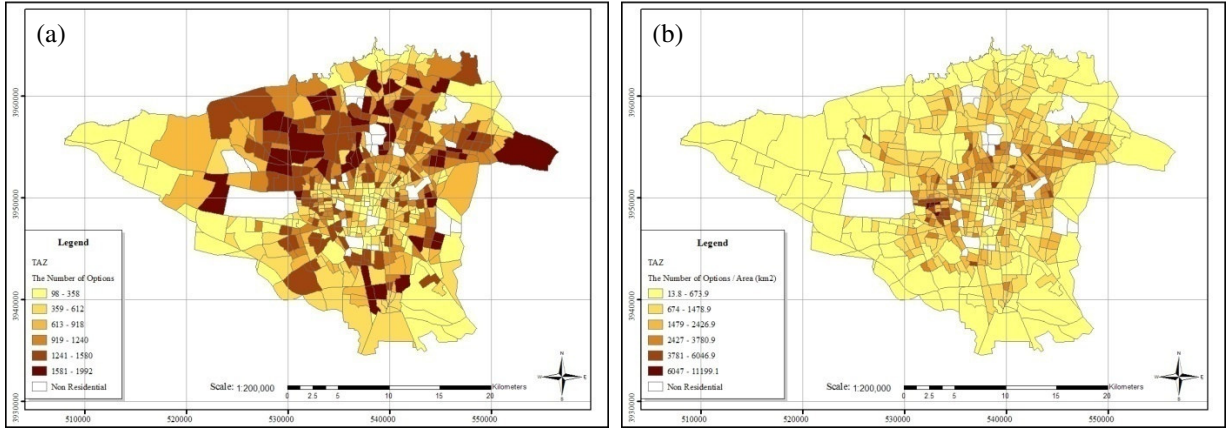
Figure 1: Spatial distribution of (a) Agents' residential options and (b) density of Agents' residential options per square kilometer

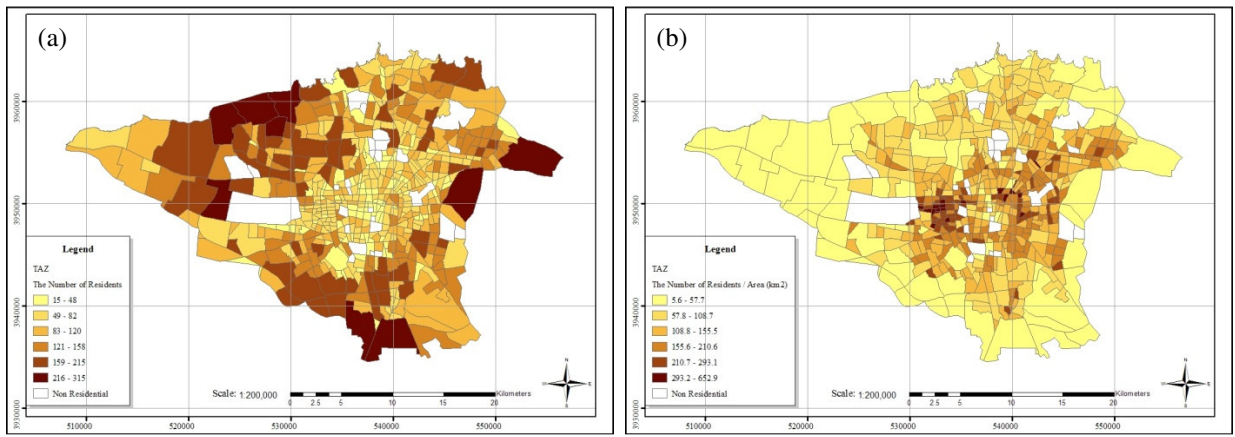
Figure 2: Spatial distribution of (a) Agents' final residence and (b) density of Agents' final residence per square kilometer

## 4. Validation

The proposed model is validated using a sample data composed of 1350 renter households derived from the survey of residential and travel preferences of households in Tehran (TCTTS, 2012). These data are not used for calibration of the proposed model in the Monte Carlo simulation. Residence of these sample households is simulated by the model and results are compared with their actual residence. Results indicate that the proposed model is able to correctly simulate the actual residential zone of 59.3% of households. Figure 3 shows the simulation accuracy of the households' residence by distance between their actual and simulated residences. As shown in this Figure, residence of 81.4% of the sample households is simulated in distances of less than 5 km from their actual residence which is an indicator of a good performance of the proposed model.

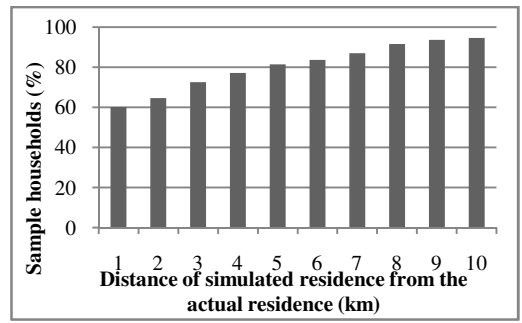
Figure 3: Simulation accuracy of residence of sample households by distance between their actual and simulated residences

## 5. Transport development scenarios

In this section, three major scenarios of the future development plan of Tehran are considered. Undoubtedly, these scenarios would have various long-term and short-term impacts on traffic, environmental pollutions, land use changes, housing prices and accessibility. But, in this study, only impacts of these scenarios on residence choice of renter households are evaluated. The considered scenarios are as below:

### *Scenario 1: A new highway construction*

In this study, effects of the southern part of Imam Ali highway on residence choice of renters are studied. As illustrated in Figure 4(a), this highway with the length of about 35 kilometers is one of the longest highways of Tehran that connects the northeast of Tehran to its southeast. Northern part of the highway has been completed and its southern part with the length of about 26 kilometers is under construction (Tehran Municipality, 2013a).

### *Scenario 2: A new subway line construction*

In this study, effects of the under construction southern part of line 3 (Figure 4(b)) with the length of about 19 kilometers and 15 stations on residence choice of renters are evaluated.

### *Scenario 3: A new BRT line construction*

In this study, effects of the under construction eastern part of line 5 (Figure 4(c)) with the length of about 9 kilometers on residence choice of renters are examined.

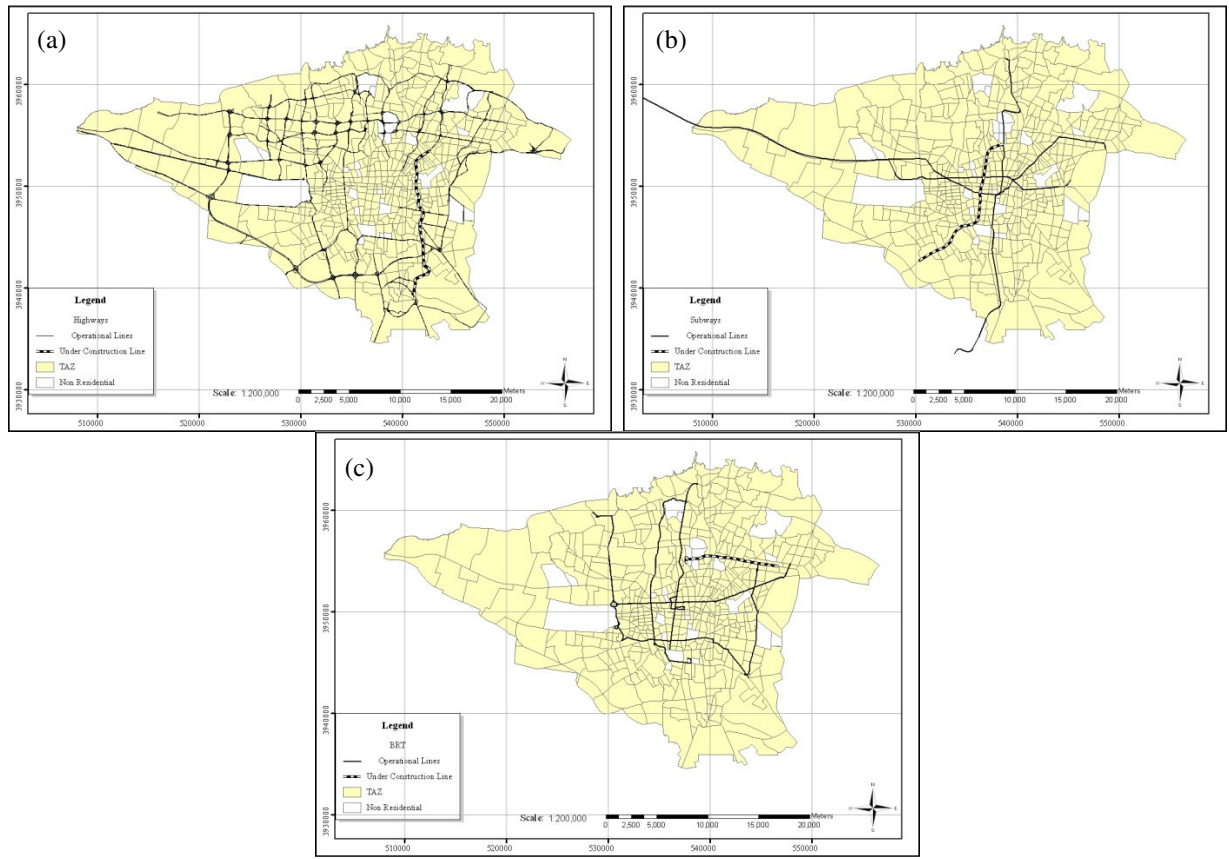

Figure 4: The existing and under construction (a) highways, (b) subway lines; and (c) BRT lines in Tehran

## 5.1. Changes in accessibility and housing rent

The transport development scenarios significantly improve the accessibility in their neighborhoods. Therefore, new accessibility values of TAZs are updated using Eq. (3). In addition, due to improvement of the accessibility, the housing rent is changed in the neighborhoods of these scenarios. Changes of housing rent are estimated by consulting with real estate agencies and experiences gained from construction of other similar transportation projects in Tehran. Investigations show that in contrast to BRT which shows no considerable effects on the housing rent, highway and subway developments have significant effects on the housing rent. Table 3 represents the rate of changes in the housing rent within different neighborhoods of the highway and subway projects. These heuristic values are used in Eq. (10) as $P_r$ to calculate the new housing rent of TAZs which are within neighborhoods of the new highway and subway.

$$\text{Eq. (10):} \qquad R_i^n = \frac{(A_i - \sum B_i^r)R_i + (\sum B_i^r)P_r R_i}{A_i}$$

where:
$R_i^n$ and $R_i$ are the new and previous housing rent per square meter in zone (*i*), respectively;
*r* is the neighborhood radius;
$A_i$ and $B_i^r$ respectively are the area of zone (*i*) and the buffer area of radius (r) inside zone (*i*);
$P_r$ is the rate of change of housing rent in the neighborhood of radius *r*.

Table 3: the rate of changes in the housing rent within different neighborhoods of the highway and subway projects

|  | Neighborhood radius (km) | Rate of change (%) |
|---|---|---|
| **Highway** | 0-0.1 | -15 |
|  | 0.1-1 | 15 |
|  | 1-1.5 | 10 |
|  | 1.5-2 | 5 |
| **Subway stations** | 0-0.5 | 15 |
|  | 0.5-1.2 | 10 |
|  | 1.2-1.9 | 5 |

## 5.2. Changes in residence choice of renter households

For studying how residential demand and residents of zones are changed by each scenario, the proposed model is rerun after implementation of each scenario and resulting changes in residential options and residence of agents are evaluated (Figures 5, 6 and 7). As indicated in Figures 5(a), 6(a) and 7(a), residential demand generally increases in neighboring zones of the new transport developments and decreases in farther zones. This illustrates that some agents have preferred to move to neighboring zones of the new transport facilities due to improvement of accessibility to transport services in these zones. In other words, neighboring zones of the new transport facilities attract some residential demands from farther zones. Also, because of high residential demand and subsequently more intense competition for residence choice in neighboring zones of the new transport facilities, considerable changes are occurred in the composition of residents in these zones (Figures 5(b), 6(b) and 7(b)).

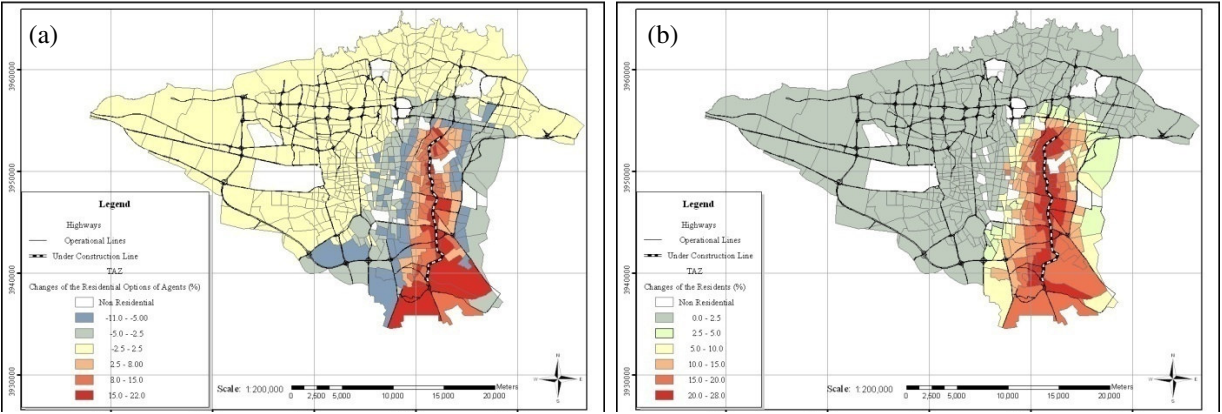

Figure 5: Changes in (a) residential options and (b) residence of agents after implementation of the new highway

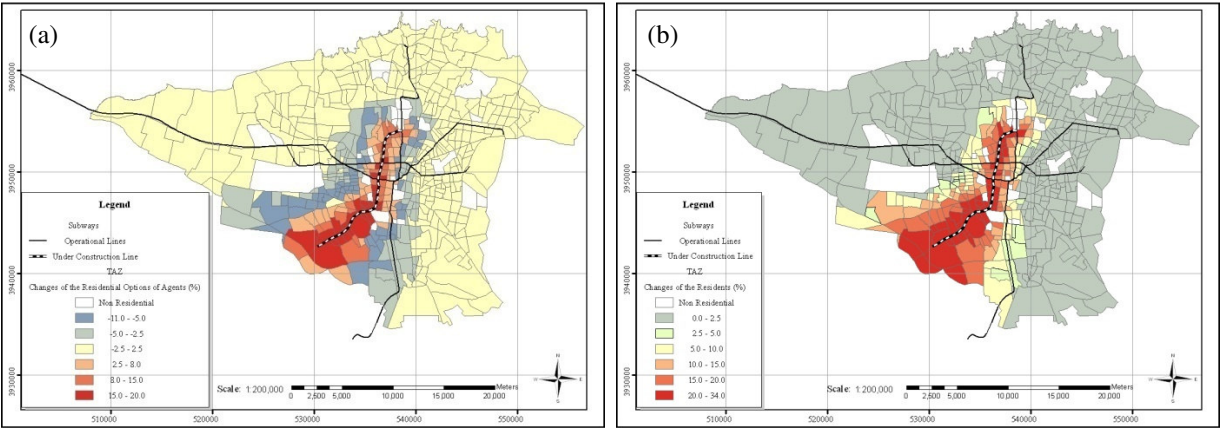

Figure 6: Changes in (a) residential options and (b) residence of agents after implementation of the new subway line

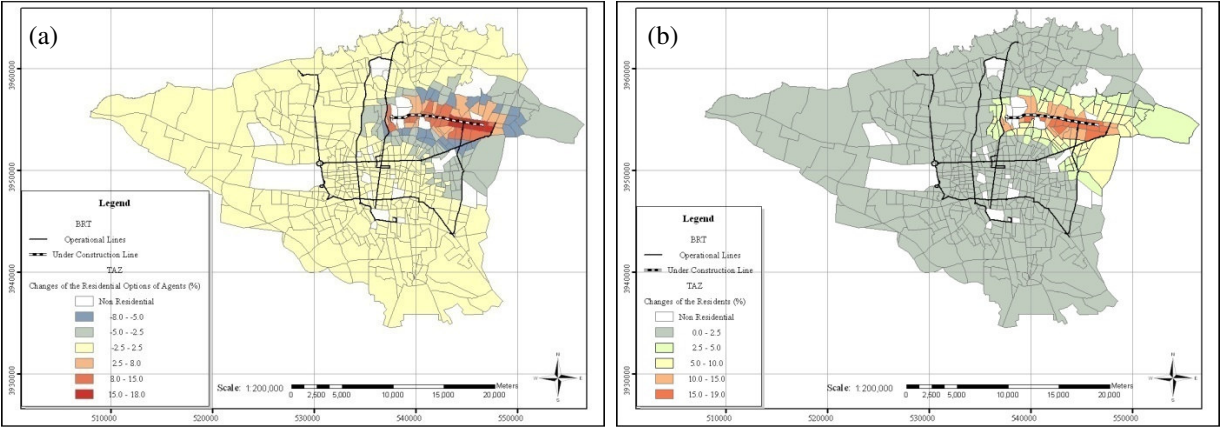

Figure 7: Changes in (a) residential options and (b) residence of agents after implementation of the new BRT line

Changes of residential demand and residents vary in different geographical directions around the new transport developments. In scenario 1, these changes are greater in southern and central neighboring zones of the new highway (Figure 5). The main reason for this may be that the accessibility of these zones to the highway network was very low before implementation of the new highway. Therefore, after implementation of this highway, residential demand significantly increases in these zones due to considerable improvements in accessibility of these zones to the highway network. In addition, because of dominant pattern of daily trips in these areas which is from south to north, the new highway may be more attractive for residents of the southern neighborhoods. As illustrated in Figure 6, the new subway line significantly improves the accessibility of many deprived areas of the southwest of Tehran to the subway network. Therefore, this line shows substantial effects on residential demand and composition of residents in these areas. In other words, because of the lack of subway stations in southwestern zones, residents of these zones pay higher attention to the new constructed subway line in their residence choice. Figure 7 shows that the new BRT line has more influences on residential demand and composition of residents in its southern and eastern neighboring zones. This suggests that residents of these zones, who generally have lower income levels, show higher interest to use the new BRT line, perhaps due to the prevailing pattern of daily trips in these areas which is from east to west and also the lower costs of traveling by BRT.

Figure 8 represents changes in the mean income level of residents in residential zones after implementation of the scenarios. As illustrated in Figures 8(a) and (b), the mean income of residents is generally increased in neighboring zones of the new highway and subway line. There are some reasons for this; first, the housing rent generally increases in these zones due to improvement of the accessibility to transport services. Second, the competition for residing in these zones is more intensive due to the increase of residential demand which consequently results in success of agents with higher incomes to reside in these zones. However, the situation is completely different in neighborhoods of the new BRT line. As shown in Figure 8(c), the mean income level decreases in neighboring zones of the new BRT line, because high-income agents do not show interests to move to these zones.

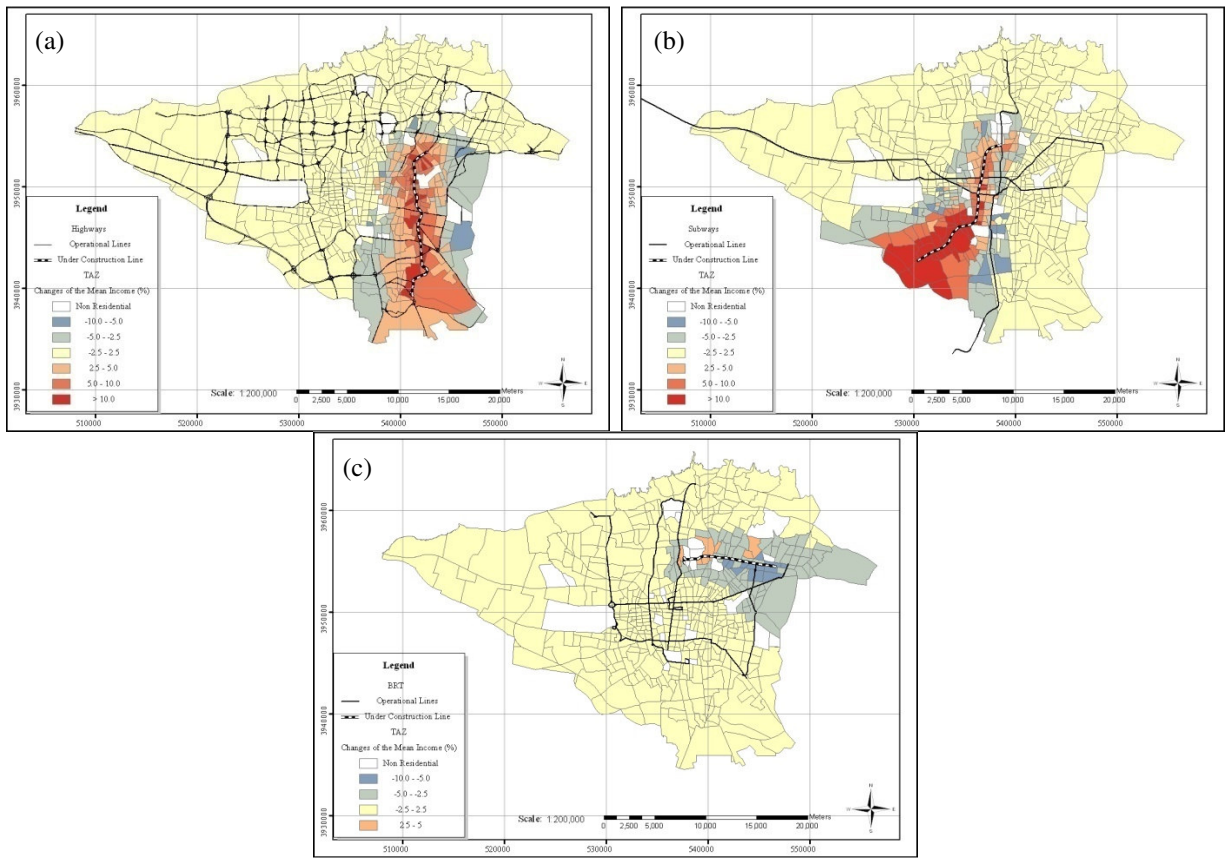

Figure 8: Changes in the mean income in TAZs after implementation of the new (a) highway, (b) subway line, and (c) BRT line

As seen in Figure 9, the mean car ownership is meaningfully changed in neighborhoods of the new transport developments. This attribute is generally increased in neighboring zones of the new highway and subway line. This indicates that agents with one or more cars are more interested to reside in these zones. However, a part of this increase may result from the increase of the mean income level in these zones, because there is a strong correlation between the income level and the number of cars owned by households. Also, it should be noted that the increase of mean car ownership in neighborhoods of the new subway line generally is lower than those of the new highway neighborhoods. The increase of mean car ownership in neighboring zones of the new subway line shows that this public transport mode attracts agents with one or more cars. Therefore, this suggests that development of the subway network may reduce the use of private car among agents having one or more cars. On the other side, the new BRT line, as planned in scenario 3, does not show meaningful effects on the mean car ownership in its neighboring residential zones. Although the more intensive competition among agents results in residing of agents with higher incomes and thereby causes a slight increase in the mean car ownership in some zones, other zones show a decreasing trend, because agents with fewer cars generally are more interested to reside in neighborhoods of the BRT network. As a result, it can be said that agents with more cars do not show great interests to move to neighborhoods of the new BRT line.

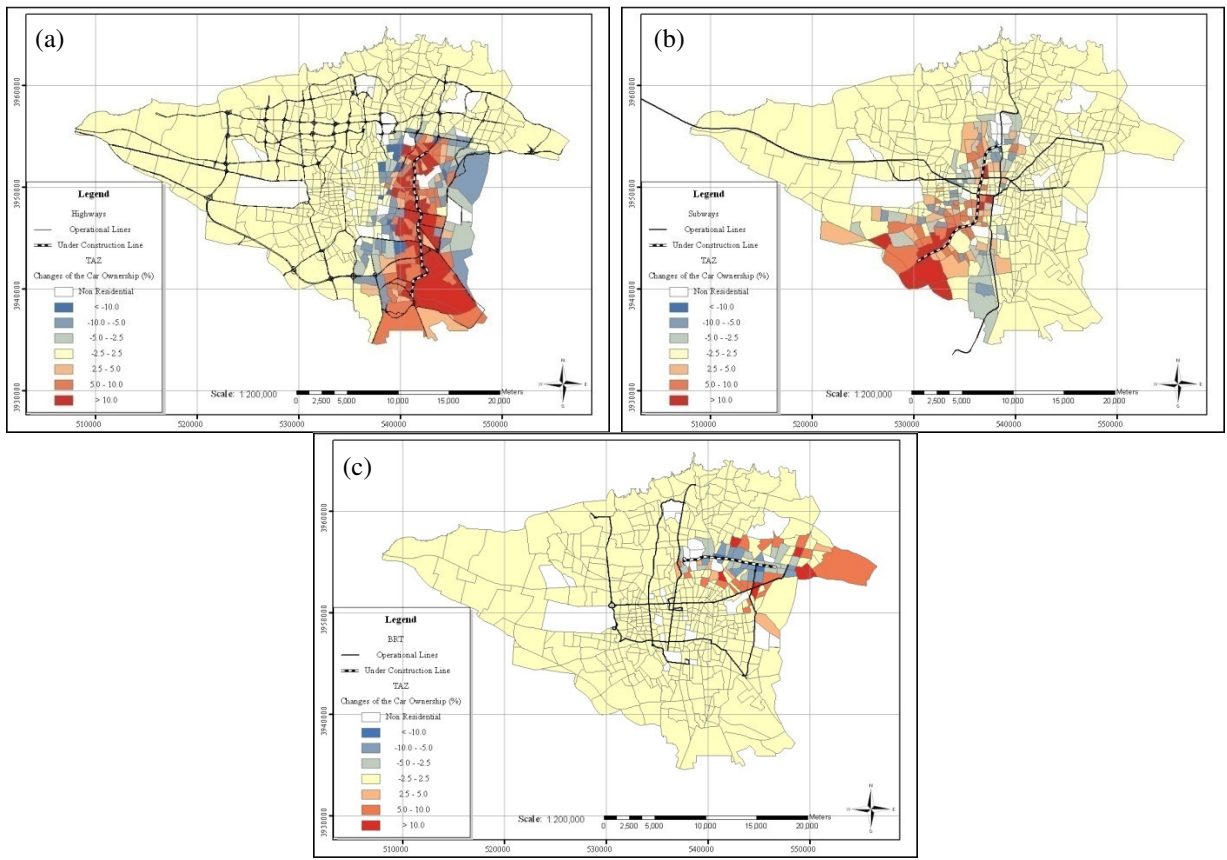

Figure 9: Changes in the mean car ownership in TAZs after implementation of the new (a) highway, (b) subway line, and (c) BRT line

The transport development scenarios result in slight increase of the number of agents who are unable to reside in any zone. The number of these agents increases from 1.36% to 1.51% and 1.58% after implementation of scenarios 1 and 2, respectively. However, the number of these agents is not changed in scenario 3. The main reason of this is that after implementation of the new highway and subway line, the housing rent increases in the southern poor areas and therefore some of the low-income agents who previously resided in these areas cannot reside again there. A summary of changes in residence choice of agents after implementation of the transport development scenarios are presented in Table 4.

Table 4: Changes in residential demand and residence of agents after implementation of the transport development scenarios

|  | Scenario 1 (%) | Scenario 2 (%) | Scenario 3 (%) |
|---|---|---|---|
| Percentage of zones in which residential demand is changed | 48.1 | 46.8 | 27.1 |
| The maximum increase in residential demand of zones | 21.4 | 19.9 | 17.8 |
| The maximum decrease in residential demand of zones | 10.9 | 10.8 | 8.0 |
| Percentage of zones which their residents are changed (affected zones) | 37.8 | 39.5 | 20.1 |
| The maximum change in residents of zones | 28.0 | 33.3 | 18.5 |
| Percentage of agents who change their residence | 6.3 | 5.6 | 1.4 |
| [1] Percentage of agents whose new and former residences are less than 2.5 km apart | 53.7 | 48.8 | 61.1 |
| [1] Percentage of agents whose new and former residence are more than 5 km apart | 4.1 | 5.5 | 2.7 |
| [2,3] Percentage of high-income agents who move to neighborhoods of the new transport services | 54.3 | 43.7 | 11.0 |
| [2,3] Percentage of low-income agents who move to neighborhoods of the new transport services | 9.4 | 18.6 | 59.5 |
| [2] Percentage of agents who own more than one car and move to neighborhoods of the new transport services | 28.6 | 16.0 | 6.1 |
| [2] Percentage of agents who have no car and move to neighborhoods of the new transport services | 7.5 | 45.1 | 62.9 |
| Changes in percentage of agents with no residence | +9.1 | +15.7 | 0 |

[1] This parameter is calculated relative to all agents who change their residence.
[2] This parameter is calculated relative to all agents who move to neighborhoods of the new transport services.
[3] The income level of agents is determined with respect to the mean income of residents in neighborhoods of the transport developments.

## 6. Conclusion

There are complex relationships between transportation and residence choice of households. By changing accessibility to various opportunities and services, transport developments may affect residence choice of different socio-economic categories of renter households. In this paper, an agent-based model was developed to study these effects. In this model, renter households were represented as agents who can individually decide, choose, compete and reside according to their residential criteria and preferences. Agent-based modeling is currently known as an efficient disaggregated approach to model large and complex socio-economic systems and processes. The main advantage of agent-based models is their ability to represent residence choice behavior of individual households.

The developed agent-based model was used to evaluate effects of three transport development scenarios including construction of a new highway, subway, and BRT line on residence choice of renter households in Tehran. These scenarios lead to considerable effects on residential demand and composition of residents in many zones such that residents of some zones are changed up to 33 percent. Various socio-economic categories of households show different residence choice behaviors in resulting situations from implementation of each scenario. High-income households owning one or more cars show the highest sensitivity to development of the highway network. Whilst in the case of BRT line development, households with low incomes and without car are more affected. It is interesting that in development of the subway network, both categories of households show some interests to live near the new subway line. As a result, it can be said that subway is attractive for various socio-economic categories of residents in Tehran. Particularly, subway is a more attractive option than BRT for developing public transportation where residents have high incomes and more than one car. In addition, after development of the new highway and subway, the number of households who do not afford to reside in any zone slightly increases. This shows significant impacts of transport development policies on residence choice of tenants which even may result in loss of home for some renter households.

However, because of using traffic analysis zones (TAZs) as the spatial units, findings of this paper are exposed to effects of the modifiable areal unit problem. These effects require close considerations by testing the model with varying sizes and configurations of spatial units. Also, in this paper, only residence choice behavior of renter households was investigated, but the model can be extended to include all types of households. In addition, the time distance can be used to more accurately measure accessibilities to various opportunities in future studies, especially when large-scale units such as census blocks or parcels are used. The proposed model can be also extended to determine the housing rent in a price bidding framework. Finally, the model can be developed to evaluate effects of different transport policies (e.g. traffic restrictions on private cars and changes in the existing transportation networks), housing policies (e.g. housing assistance programs to low-income households, changes in the housing supply in some areas and changes in housing taxes), and land use policies (e.g. changes in distribution of workplaces and public facilities) on residence choice behavior of various households.